\def\mydate{\date{Feb. 14, 2006}}

\documentclass[aps,prl,twocolumn,superscriptaddress,showpacs]{revtex4}

\def\figonescale{0.4}
\def\figtwoscale{0.49}
\def\bellelogo{\vbox to 16mm{
               \vss\hbox to \textwidth{\resizebox{!}{2cm}{
               \includegraphics{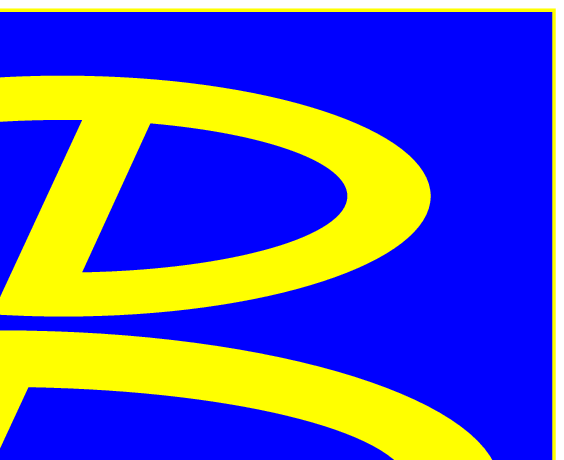}}\hss }}\vspace{-1cm}}
\def\preprintA{\hbox{\hfil KEK Preprint 2005-101}}
\def\preprintB{\hbox{\hfil Belle Preprint 2006-5}}

\usepackage{graphicx}

\def\KM{K^-}
\def\KS{{K^0_S}}

\def\piZ{{\pi^0}}
\def\piP{{\pi^+}}
\def\piM{{\pi^-}}
\def\rhoM{{\rho^-}}
\def\rhoZ{{\rho^0}}
\def\Kstar{{K^*}}
\def\KstarM{{K^{*-}}}
\def\KstarB{{\Kbar^{*0}}}
\def\omegaG{{\omega\gamma}}

\def\rhoMG{{\rhoM\gamma}}
\def\rhoZG{{\rhoZ\gamma}}

\def\KstarG{{\Kstar\gamma}}
\def\KstarBG{{\KstarB\gamma}}
\def\KstarMG{{\KstarM\gamma}}
\def\qqbar{q\overline{q}{}}
\def\Bbar{\overline{B}{}}
\def\Kbar{\overline{K}{}}
\def\epem{e^+e^-{}}

\def\btosgamma{b\to s\gamma}
\def\btodgamma{b\to d\gamma}

\def\BtoRG{\Bbar\to \rho\gamma}
\def\BtoRMG{B^-\to \rho^-\gamma}
\def\BtoRZG{\Bbar^0\to \rho^0\gamma}
\def\BtoROG{\Bbar\to (\rho,\omega)\gamma}
\def\BtoOMG{\Bbar^0\to \omega\gamma}
\def\BtoOG{\Bbar^0\to \omega\gamma}

\def\BtoKG{\Bbar\to \Kbar^*\gamma}
\def\BtoKBG{\Bbar^0\to \Kbar^{*0}\gamma}
\def\BtoKMG{B^-\to K^{*-}\gamma}

\def\BtoXsgamma{B\to X_s\gamma}

\def\cm{\mbox{~cm}}

\def\GeV{\mbox{~GeV}}
\def\GeVc{\mbox{~GeV}/c}
\def\GeVcc{\mbox{~GeV}/c^2}

\def\MeVc{\mbox{~MeV}/c}
\def\MeVcc{\mbox{~MeV}/c^2}

\def\Vcb{V_{cb}}
\def\Vtd{V_{td}}
\def\Vts{V_{ts}}

\def\Br{{\cal B}}

\def\Lpi{{\cal L}_\pi}
\def\LK{{\cal L}_K}

\def\Mbc{M_{\rm bc}}
\def\DeltaE{\Delta{E}}
\def\Ebeam{E^*_{\rm beam}{}}

\def\MKpi{M_{K\pi}}

\def\Egamma{E_\gamma}

\def\tauBratio{{\tau_{B^+}\over\tau_{B^0}}}

\def\piZeta{\piZ/\eta}

\def\thetaB{{\theta^*_B}}
\def\cosB{{\cos\thetaB}}

\def\calF{{\cal F}}

\def\calR{{\cal R}}

\def\Deltaz{\Delta{z}}
\def\thetahel{\theta_{\rm hel}}
\def\coshel{\cos\thetahel}

\def\Lzero{{\cal L}_0}
\def\Lmax{{\cal L}_{\rm max}}

\def\PM#1#2{\,^{+#1}_{-#2}{}}
\def\EM#1{\times10^{-#1}}

\def\etal{\textit{et al.}}
\def\Journal#1#2#3#4{{#1} {\bf #2}, #3 (#4)} 
\def\NIMA{Nucl. Instrum. Meth. A}
\def\NPB{Nucl. Phys. B}
\def\PLB{Phys. Lett. B}
\def\PRL{Phys. Rev. Lett.}
\def\PRD{Phys. Rev. D}
\def\ZPC{Z. Phys. C}
\def\EPJC{Eur. Phys. J. C}
\def\JPG{J. Phys. G}

\makeatletter
\def\myep@[#1]#2{\resizebox{#1\textwidth}{!}{\includegraphics{#2}}}
\def\myeps{\@ifnextchar[{\myep@}{\myep@[1]}}
\makeatother

\def\drcut{0.5\cm} 
\def\dzcut{3.0\cm} 
\def\ptrkcut{100\MeVc}

\def\EffKaon{90\%}
\def\EffPionRM{83\%}
\def\EffPionRZ{81\%}
\def\EffPionOM{91\%}
\def\FakePionRM{5.8\%}
\def\FakePionRZ{6.3\%}
\def\FakePionOM{8.4\%}

\def\BrBtoKG{41.1\PM{ 1.4}{ 1.3}}                
\def\signifROGstat{5.4}                          
\def\BrBtoROG{1.32\PM{0.34}{0.31}\PM{0.10}{0.09}} 

\def\NSBtoRMG{8.5}                               
\def\signifRMGstat{1.6}                          
\def\BrBtoRMG{0.55\PM{0.42}{0.36}\PM{0.09}{0.08}} 

\def\NSBtoRZG{20.7}                              
\def\signifRZGstat{5.2}                          
\def\BrBtoRZG{1.25\PM{0.37}{0.33}\PM{0.07}{0.06}} 

\def\NSBtoOMG{5.7}                               
\def\signifOMGstat{2.6}                          
\def\BrBtoOMG{0.56\PM{0.34}{0.27}\PM{0.05}{0.10}} 

\def\EffRM{3.86\pm0.23}
\def\EffRZ{4.30\pm0.28}
\def\EffOM{2.61\pm0.21}

\def\signifRMconv{1.6}  
\def\signifRZconv{5.2}  
\def\signifRMconv{1.6}  
\def\signifRZconv{5.2}  
\def\signifOMconv{2.3}  
\def\signifROratioconv{5.1}  

\def\NSBtoROG{36.9}  

\def\BrBtoROGfull{1.32\PM{0.34}{0.31}{\rm(stat.)}\PM{0.10}{0.09}{\rm(syst.)}}
\def\ratioBtoROGfull{0.032\pm0.008{\rm(stat.)}\pm0.002{\rm(syst.)}}

\def\VtdoverVtsValFull{0.199\PM{0.026}{0.025}{\rm(exp.)}\PM{0.018}{0.015}{\rm(theo.)}}

\def\VtdoverVtsLo{0.142} 
\def\VtdoverVtsHi{0.259} 

\begin{document}


\title{
  \vbox{\bellelogo \vspace{-1cm}
  \hbox to \textwidth{\rm\normalsize\hss\preprintA}
  \hbox to \textwidth{\rm\normalsize\hss\preprintB}}
  \vspace*{15mm}Observation of \boldmath$\btodgamma$ and
       Determination of $|\Vtd/\Vts|$}


\affiliation{Budker Institute of Nuclear Physics, Novosibirsk}
\affiliation{Chonnam National University, Kwangju}
\affiliation{University of Cincinnati, Cincinnati, Ohio 45221}
\affiliation{University of Frankfurt, Frankfurt}
\affiliation{University of Hawaii, Honolulu, Hawaii 96822}
\affiliation{High Energy Accelerator Research Organization (KEK), Tsukuba}
\affiliation{Institute of High Energy Physics, Chinese Academy of Sciences, Beijing}
\affiliation{Institute of High Energy Physics, Vienna}
\affiliation{Institute of High Energy Physics, Protvino}
\affiliation{Institute for Theoretical and Experimental Physics, Moscow}
\affiliation{J. Stefan Institute, Ljubljana}
\affiliation{Kanagawa University, Yokohama}
\affiliation{Korea University, Seoul}
\affiliation{Kyungpook National University, Taegu}
\affiliation{Swiss Federal Institute of Technology of Lausanne, EPFL, Lausanne}
\affiliation{University of Ljubljana, Ljubljana}
\affiliation{University of Maribor, Maribor}
\affiliation{University of Melbourne, Victoria}
\affiliation{Nagoya University, Nagoya}
\affiliation{Nara Women's University, Nara}
\affiliation{National Central University, Chung-li}
\affiliation{National United University, Miao Li}
\affiliation{Department of Physics, National Taiwan University, Taipei}
\affiliation{H. Niewodniczanski Institute of Nuclear Physics, Krakow}
\affiliation{Nippon Dental University, Niigata}
\affiliation{Niigata University, Niigata}
\affiliation{Nova Gorica Polytechnic, Nova Gorica}
\affiliation{Osaka City University, Osaka}
\affiliation{Osaka University, Osaka}
\affiliation{Panjab University, Chandigarh}
\affiliation{Peking University, Beijing}
\affiliation{Princeton University, Princeton, New Jersey 08544}
\affiliation{RIKEN BNL Research Center, Upton, New York 11973}
\affiliation{Saga University, Saga}
\affiliation{University of Science and Technology of China, Hefei}
\affiliation{Seoul National University, Seoul}
\affiliation{Sungkyunkwan University, Suwon}
\affiliation{University of Sydney, Sydney NSW}
\affiliation{Tata Institute of Fundamental Research, Bombay}
\affiliation{Toho University, Funabashi}
\affiliation{Tohoku Gakuin University, Tagajo}
\affiliation{Tohoku University, Sendai}
\affiliation{Department of Physics, University of Tokyo, Tokyo}
\affiliation{Tokyo Institute of Technology, Tokyo}
\affiliation{Tokyo Metropolitan University, Tokyo}
\affiliation{Tokyo University of Agriculture and Technology, Tokyo}
\affiliation{University of Tsukuba, Tsukuba}
\affiliation{Virginia Polytechnic Institute and State University, Blacksburg, Virginia 24061}
\affiliation{Yonsei University, Seoul}
   \author{D.~Mohapatra}\affiliation{Virginia Polytechnic Institute and State University, Blacksburg, Virginia 24061} 
   \author{M.~Nakao}\affiliation{High Energy Accelerator Research Organization (KEK), Tsukuba} 
   \author{S.~Nishida}\affiliation{High Energy Accelerator Research Organization (KEK), Tsukuba} 
   \author{K.~Abe}\affiliation{High Energy Accelerator Research Organization (KEK), Tsukuba} 
   \author{K.~Abe}\affiliation{Tohoku Gakuin University, Tagajo} 
   \author{I.~Adachi}\affiliation{High Energy Accelerator Research Organization (KEK), Tsukuba} 
   \author{H.~Aihara}\affiliation{Department of Physics, University of Tokyo, Tokyo} 
   \author{D.~Anipko}\affiliation{Budker Institute of Nuclear Physics, Novosibirsk} 
   \author{K.~Arinstein}\affiliation{Budker Institute of Nuclear Physics, Novosibirsk} 
   \author{Y.~Asano}\affiliation{University of Tsukuba, Tsukuba} 
% ...
\collaboration{The Belle Collaboration}



\mydate

\begin{abstract} 

  We report the observation of the flavor-changing neutral current
  process $\btodgamma$ using a sample of $386\times10^6$ $B$ meson pairs
  accumulated by the Belle detector at the KEKB $\epem$ collider.  We
  measure branching fractions for the exclusive modes $\BtoRMG$,
  $\BtoRZG$ and $\BtoOMG$.  Assuming that these three modes are related
  by isospin, we find $\Br(\BtoROG)=(\BrBtoROGfull)\EM6$ with a
  significance of $\signifROratioconv\sigma$.  This result is used to
  determine the ratio of CKM matrix elements $|\Vtd/\Vts|$ to be
  $\VtdoverVtsValFull$.

\end{abstract}


\pacs{11.30.Hv, 13.40.Hq, 14.65.Fy, 14.40.Nd}

\maketitle



\begin{figure}[ht]
\begin{center}
\vspace*{24pt}
\myeps[\figonescale]{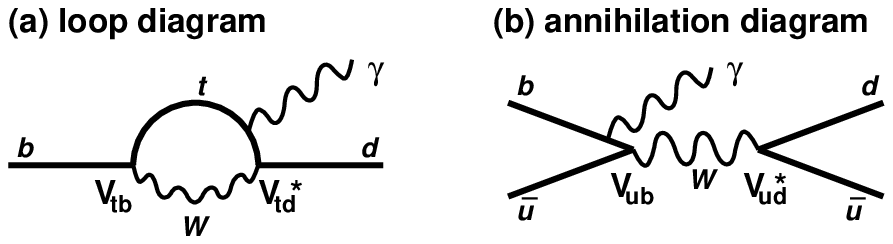}
\caption{(a) Loop diagram for $\btodgamma$ and (b) annihilation diagram,
  which contributes only to $\BtoRMG$.}
\label{fig:diagram}
\end{center}
\end{figure}

The $\btodgamma$ process, which proceeds via a loop diagram
(Fig.~\ref{fig:diagram}(a)) in the Standard Model (SM), is suppressed
with respect to $\btosgamma$ by the Cabibbo-Kobayashi-Maskawa (CKM)
factor~\cite{bib:ckm} $|\Vtd/\Vts|^2 \sim 0.04$, with large uncertainty
due to the lack of precise knowledge of $|\Vtd|$.  The exclusive modes
$\BtoRG$ and $\BtoOG$ are presumably the easiest modes to search for;
no evidence for the decays has been previously
reported~\cite{bib:belle-rhogam,bib:babar-rhogam}.
%
%
The predicted branching fractions are
{$(0.9\mbox{--}2.7)\EM6$}~\cite{bib:ali-cdlu,bib:bosch-buchalla}
based on the measured rate for the $\btosgamma$ process $\BtoKG$ and the
$|\Vtd/\Vts|^2$ factor with corrections due to form factors, $SU(3)$
breaking effects, and, for the $B^-$ decay, inclusion of an annihilation
diagram (Fig.~\ref{fig:diagram}(b)).  Measurement of these exclusive
branching fractions allows one to determine the value of $|\Vtd/\Vts|$
in the context of the SM and to search for physics beyond the
SM~\cite{bib:rhogam-bsm}.  In this Letter, we report the observation of
the $\btodgamma$ process using a sample of $(386\pm5)\times10^6$ $B$
meson pairs accumulated at the $\Upsilon(4S)$ resonance.  With a larger
data sample and an improved analysis procedure, the results supersede
those of our previous publication~\cite{bib:belle-rhogam}.


The data are produced in $\epem$ annihilation at the KEKB
energy-asymmetric (3.5 on 8 GeV) collider~\cite{bib:kekb} and collected
with the Belle detector~\cite{bib:belle-detector}, which includes a
silicon vertex detector (SVD), a central drift chamber (CDC), aerogel
threshold Cherenkov counters (ACC), time-of-flight (TOF) scintillation
counters, and an electromagnetic calorimeter (ECL) of CsI(Tl) crystals
located inside a 1.5 T superconducting solenoid coil.



We reconstruct three signal modes, $\BtoRMG$, $\BtoRZG$ and $\BtoOG$,
and two control samples, $\BtoKMG$ and $\BtoKBG$.  Charge conjugate
modes are implicitly included throughout this Letter.  The following
decay modes are used to reconstruct the intermediate states:
$\rhoM\to\piM\piZ$, $\rhoZ\to\piP\piM$, $\omega\to\piP\piM\piZ$,
$\KstarM\to\KM\piZ$, $\KstarB\to\KM\piP$, and $\piZ\to\gamma\gamma$.

Photon candidates are reconstructed from ECL energy clusters with a
photon-like shape and no associated charged track.  A photon in the
barrel ECL ($33^\circ<\theta_\gamma<128^\circ$ in the laboratory frame
polar angle) with a center-of-mass (c.m.)\ energy in the range
$1.8\GeV<\Egamma<3.4\GeV$ is selected as the primary photon candidate.
To suppress backgrounds from $\piZ/\eta\to\gamma\gamma$ decays, we apply
a veto algorithm based on likelihoods to be and not to be a $\piZ/\eta$.
The likelihoods are calculated for every combination of the primary
photon and another photon in the event using the energy of the other
photon and the invariant mass of the pair.
We also reject the primary photon candidate if the ratio of the energy
in the central $3\times3$ ECL cells to that in the central $5\times5$
cells is less than 0.95.

Neutral pions are formed from photon pairs with invariant masses within
$\pm 16\MeVcc$ (${\sim}3\sigma$) of the $\piZ$ mass.  The photon momenta
are then recalculated with a $\piZ$ mass constraint.  We require the
energy of each photon to be greater than 50 (100) MeV inside (outside)
the barrel ECL.  We also require the cosine of the angle between the two
photons in the laboratory frame to be greater than 0.7; this requirement
suppresses the copious combinatorial background with momenta below
$0.6\GeVc$.

Charged pions and kaons are selected from tracks in the CDC and SVD.
Each track is required to have a transverse momentum greater than
$\ptrkcut$ and a distance of closest approach to the interaction point
of less than $\drcut$ in radius and $\pm\dzcut$ along the $z$-axis,
which is parallel to the positron beam.  We do not use a track to form
the signal candidate if, when it is combined with an oppositely charged
track, the resulting pair has an invariant mass within $\pm30\MeVcc$ of
the $\KS$ mass and a displaced vertex that is consistent with that of a
$\KS$.  We determine pion ($\Lpi$) and kaon ($\LK$) likelihoods from
ACC, CDC and TOF information and form a likelihood ratio $\Lpi/(\Lpi +
\LK)$ to separate pions from kaons.  
The criteria for pions have efficiencies of $\EffPionRM$, $\EffPionRZ$
and $\EffPionOM$ for $\rhoM$, $\rhoZ$ and $\omega$, respectively; the
corresponding kaon misidentification rates are $\FakePionRM$,
$\FakePionRZ$ and $\FakePionOM$.  For $\Kstar$ candidates, we select
kaons with an efficiency of $\EffKaon$.

Invariant masses for the $\rho$, $\omega$ and $\Kstar$ candidates are
required to be within windows of $\pm150$, $\pm30$ and $\pm75\MeVcc$,
respectively, around their nominal values.

\def\pvecB{\vec{p}{}^{\;*}_B}

Candidate $B$ mesons are reconstructed by combining a $\rho$ or $\omega$
candidate with the primary photon and calculating two variables: the
beam-energy constrained mass $\Mbc = \sqrt{ (\Ebeam/c^2)^2 -
|\pvecB/c|^{2}}$, and the energy difference $\Delta E = E^*_{B} -
\Ebeam$.  Here, $\pvecB$ and $E^*_B$ are the {c.m.}\ momentum and energy of
the $B$ candidate, and $\Ebeam$ is the {c.m.}\ beam energy.  To improve
resolution, the magnitude of the photon momentum is replaced by $(\Ebeam
- E_{\rho/\omega}^*)/c$ when the momentum $\pvecB$ is calculated.

To optimize the event selection, we study Monte Carlo (MC) events in a
signal box defined as 
$5.273\,\mathrm{GeV}/c^2<\Mbc<5.285\,\mathrm{GeV}/c^2$ and
$-0.10\,\mathrm{GeV}<\DeltaE<0.08\,\mathrm{GeV}$.
We choose selection criteria to maximize
$N_S/\sqrt{N_B}$, where $N_S$ and $N_B$ are the expected signal and the
sum of the background yields.


The dominant background arises from continuum events
($\epem\to\qqbar(\gamma)$, $q=u,d,s,c$), where a random
combination of a $\rho$ or $\omega$ candidate with a photon forms a $B$
candidate.  We suppress this background using the following quantities:
%
(1)
$\calF$, a Fisher discriminant constructed from 16 modified Fox-Wolfram
moments~\cite{bib:belle-pi0pi0,bib:fox-wolfram} and the scalar sum of
the transverse momenta of all charged tracks and photons.
(2)
$\cosB$, where $\thetaB$ is the {c.m.}\ polar angle of the $B$
candidate direction: true $B$ mesons follow a $1-\cos^2\thetaB$
distribution, while candidates in the continuum background are
almost uniformly distributed.
(3)
$\Delta z$, the separation along the $z$-axis between the decay vertex
of the candidate $B$ meson and the fitted vertex of the remaining tracks
in the event.
%
%
Discrimination is provided due to the displacement of the signal $B$
decay vertex from the other $B$, as tracks from continuum events
typically have a common vertex.
%
%
%
For each of the quantities $\calF$, $\cosB$ and $\Deltaz$, we construct
likelihood distributions for signal and continuum events.  The $\calF$,
$\cosB$ and signal $\Deltaz$ distributions are determined from MC
samples; the continuum $\Delta z$ distribution is determined from the
data sideband $5.20\GeVcc<\Mbc<5.24\GeVcc$, $-0.1\GeV<\DeltaE<0.5\GeV$.

We form product likelihoods ${\cal L}_s$ and ${\cal L}_c$ for signal and
continuum background, respectively, from the likelihood distributions
for $\calF$, $\cosB$ and (where available) $\Deltaz$.  In addition, we
use a tagging quality variable $r$ that indicates the level of
confidence in the $B$-flavor determination as described in
Ref.~\cite{bib:hamlet}.  In the $(r,\calR)$ plane defined by the tagging
quality $r$ and the likelihood ratio $\calR={\cal L}_s/({\cal L}_s+{\cal
L}_c)$, signal tends to populate the edges at $r=1$ and $\calR=1$, while
continuum preferentially populates the edges at $r=0$ and $\calR=0$.  We
divide the events into six bins of $r$ (two bins between 0 and 0.5, and
four between 0.5 and 1) and determine the minimum $\calR$ requirement
for each bin.  In the $\rhoMG$ mode, we also assign events to
the bin $0\le r<0.25$ if the tagging-side flavor is the same as the
signal-side.  The signal efficiency is ${\sim}40\%$, and ${\sim}95\%$ of
continuum background is rejected.
For the $\KstarMG$ ($\KstarBG$) mode we use the selection criteria for the
$\rhoMG$ ($\rhoZG$) mode.

We consider the following backgrounds from $B$ decays: $\BtoKG$, other
$B\to X_s\gamma$ processes, decays with a $\piZeta$ ($B\to\rho\piZ$,
$\omega\piZ$, $\rho\eta$ and $\omega\eta$), other charmless hadronic
$B$ decays, and $b\to c$ decay modes.  We find the $b\to c$ background
to be negligible.  The $\BtoKG$ background can mimic the $\BtoRG$ signal
if the kaon from the $K^*$ is misidentified as a pion. To suppress $\BtoKG$
events we calculate $\MKpi$, where the kaon mass is assigned to one of
the charged pion candidates, and reject the candidate if $\MKpi<0.95$
($0.92$)$\GeVcc$ for the $\rhoZG$ ($\rhoMG$) mode. This requirement
removes 82\% (64\%) of the $\KstarG$ background while retaining 63\%
(87\%) of the signal.  The decay chain $\BtoKBG$,
$\Kbar^{*0}\to\KS\piZ$, $\KS\to\piP\piM$ has a small contribution to
$\BtoOG$ due to the tail of the $K^*$ Breit-Wigner lineshape. In
addition, $\BtoKG$ and other $\BtoXsgamma$ decays contribute to the
background when the $\rho$ and $\omega$ candidates are formed from
random combinations of particles.

Hadronic decays with a $\piZeta$ can mimic the signal if a photon from
the $\piZ$ or $\eta\to\gamma\gamma$ decay is soft and passes the
$\piZeta$ veto.  To suppress this background, we reject the candidate if
$|\coshel|>0.75$, $0.70$ and $0.80$ for the $\rhoMG$, $\rhoZG$ and
$\omegaG$ modes, respectively, where the helicity angle $\thetahel$ is
the angle between the $\piM$ track (normal to the $\omega$ decay plane)
and the $B$ momentum vector in the $\rho$ ($\omega$) rest frame
(similarly for the $\KstarG$ modes).  Other hadronic decays make smaller
contributions.



The reconstruction efficiency for each mode is defined as the fraction
of the signal remaining after all selection criteria are applied, where
the signal yield is determined from a fit to the sum of the signal and
continuum MC samples using the procedure described below.  The total
efficiencies are listed in Table~\ref{tbl:results}.
The systematic error on the efficiency is the quadratic sum of the
following contributions, estimated using control samples: the
uncertainty in the photon detection efficiency (2.2\%) as measured in
radiative Bhabha events; charged tracking efficiency (1.0\% per track)
from partially reconstructed $D^{*+}\to D^0\piP$, $D^0\to\KS\piP\piM$,
$\KS\to\piP(\piM)$; charged pion and kaon identification
(0.7--1.7\% per track) and misidentification (15--17\%) from
$D^{*+}\to D^0\piP$, $D^0\to\KM\piP$; neutral pion detection (4.6\%)
from $\eta$ decays to $\gamma\gamma$, $\piP\piM\piZ$ and $3\piZ$;
$\calR$-$r$ and $\piZeta$ veto requirements (2.8--5.5\%) from $B^-\to
D^0\pi^-$, $D^0\to\KM\piP$ and $\Bbar^0\to D^+\pi^-$, $D^+\to
K^-\pi^+\pi^+$; the $\omega\to\piP\piM\piZ$ branching fraction (0.8\%);
and uncertainty due to MC statistics (0.5--0.7\%).


We perform an unbinned extended maximum likelihood fit to candidates
satisfying $|\DeltaE|<0.5\GeV$ and $\Mbc>5.2\GeVcc$, individually and
simultaneously for the three signal modes.  In the latter case we assume
isospin symmetry, and we also simultaneously fit the two $\BtoKG$ modes.
We describe the events in the fit region using a sum of functions for
the signal, continuum, $\KstarG$ (for the three signal modes only), and
other background hypotheses.  The signal distribution is modeled as the
product of a Crystal Ball lineshape~\cite{bib:cbls} in $\DeltaE$ to
reproduce the asymmetric ECL energy response, and a Gaussian (another
Crystal Ball lineshape) in $\Mbc$ for the mode without (with) a $\piZ$
in the final state.  The signal parameters for $\Mbc$ and $\DeltaE$ are
determined from separate fits to the $\BtoKMG$ and $\BtoKBG$ samples for
the modes with and without a neutral pion, respectively.  The branching
fraction is the only parameter that is allowed to float for the signal
component.
The continuum background component is modeled as the product of a linear
function in $\DeltaE$ and an ARGUS function~\cite{bib:argus-function}
in $\Mbc$.  
The continuum shape parameters and normalizations are mode dependent and
allowed to float.  We use the distributions of MC events to model the
shapes of other background components.
The size of the $\KstarG$ background component in
each signal mode is constrained using the fit to the $\KstarG$ events
and the known misidentification probability.  Other radiative and
charmless decays are considered as an additional background component
when we extract the signal yield.  The levels of the other backgrounds
are fixed using known branching fractions or upper
limits~\cite{bib:hfag2005}.  
We constrain branching fractions in the simultaneous fit using the
isospin relations~\cite{bib:ali-1994,bib:ali-cdlu}
$\Br(\BtoROG)
\equiv
\Br(\BtoRMG) = 2\tauBratio\Br(\BtoRZG) = 2\tauBratio\Br(\BtoOG)$ and
$\Br(\BtoKG) \equiv
\Br(\BtoKMG) = \tauBratio\Br(\BtoKBG)$, where $\tauBratio =
1.076\pm0.008$~\cite{bib:hfag2005}.


\begin{figure}[ht]
\begin{center}
\myeps[\figtwoscale]{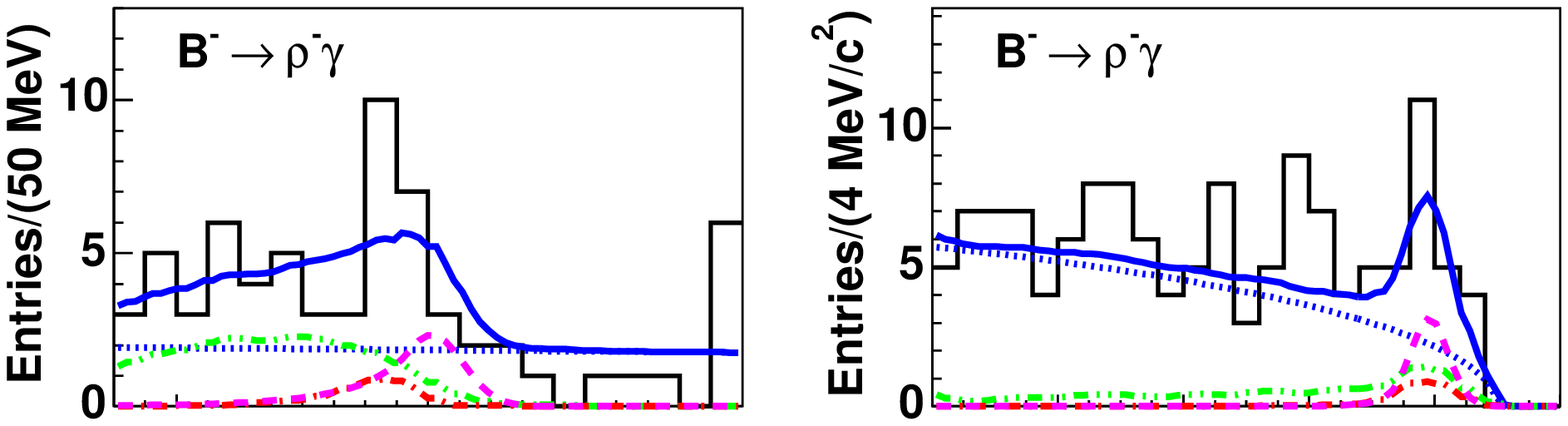}\\
\myeps[\figtwoscale]{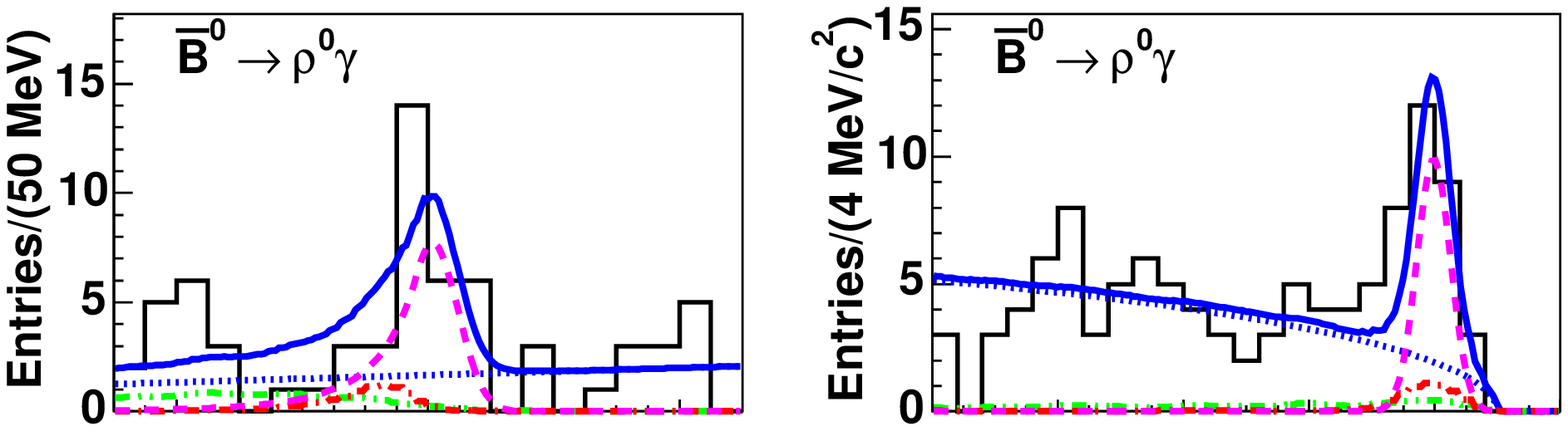}\\
\myeps[\figtwoscale]{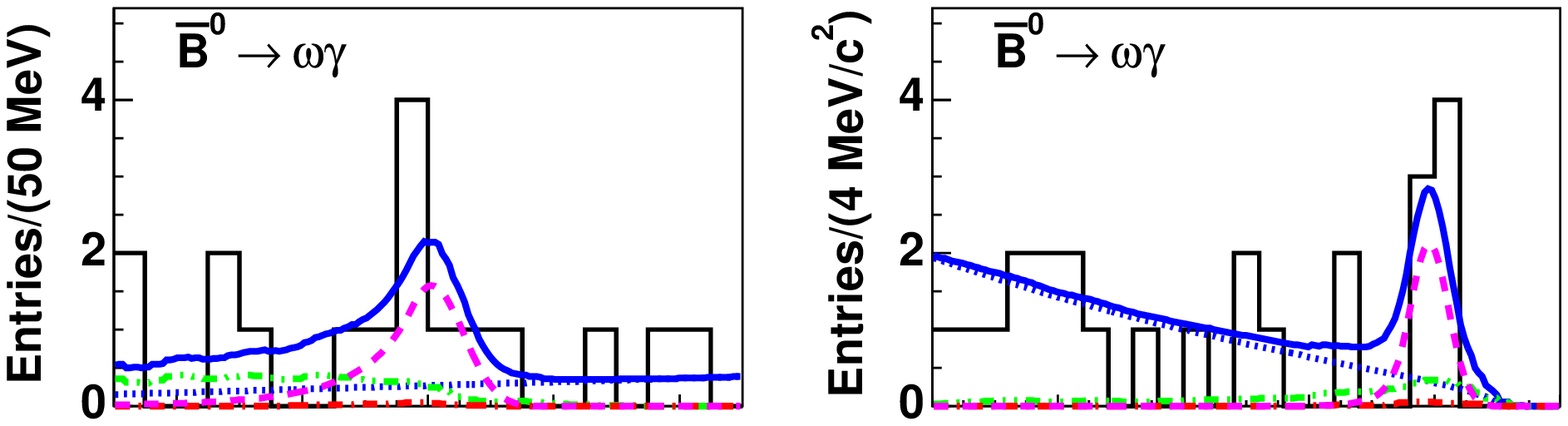}\\
\myeps[\figtwoscale]{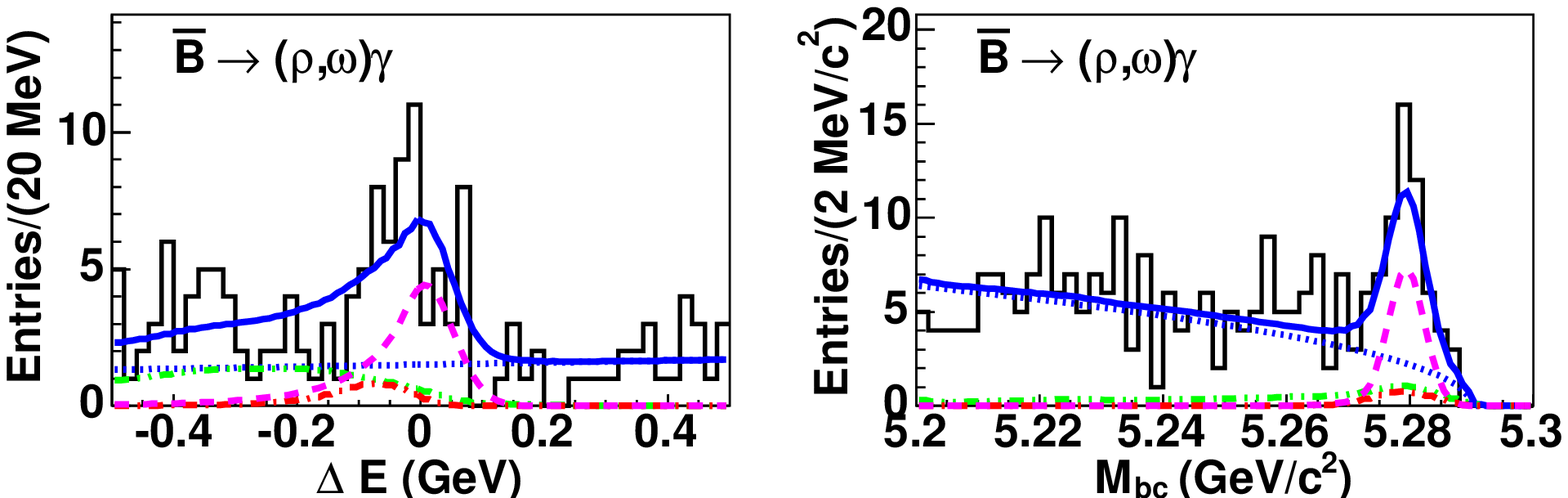}
\caption{Projections of the fit results to $\Mbc$ (in the
         region $-0.10\GeV<\DeltaE<0.08\GeV$) and $\DeltaE$ (in the
         region $5.273\GeVcc<\Mbc<5.285\GeVcc$) for the individual and
         simultaneous fits.
         Curves show the signal (dashed), continuum (dotted),
         $\BtoKG$ (dot-dashed), other $B$ decay
         background (dot-dot-dashed) components, and the total fit result
         (solid).}
\label{fig:simfit}
\end{center}
\end{figure}


\begin{table}[ht]
\caption{ Yield, significance with (without) systematic
  uncertainty, efficiency, and branching fraction ($\Br$) for each mode. }
\label{tbl:results}
\begin{ruledtabular}
\begin{tabular}{lcccc}
Mode & Yield & Signif. & Efficiency (\%) & $\Br$ ($10^{-6}$) \\
\hline
$\BtoRMG$ & $\NSBtoRMG$ & $\signifRMconv$ ($\signifRMGstat$) & $\EffRM$ & $\BrBtoRMG$ \\
$\BtoRZG$ & $\NSBtoRZG$ & $\signifRZconv$ ($\signifRZGstat$) & $\EffRZ$ & $\BrBtoRZG$ \\
$\BtoOMG$ & $\NSBtoOMG$ & $\signifOMconv$ ($\signifOMGstat$) & $\EffOM$ & $\BrBtoOMG$ \\
$\BtoROG$ & $\NSBtoROG$ & $\signifROratioconv$ ($\signifROGstat$) & --- & $\BrBtoROG$ \\
%
\end{tabular}
\end{ruledtabular}
\end{table}

The results of the fits are shown in Fig.~\ref{fig:simfit} and listed in
Table~\ref{tbl:results}.
The simultaneous fit gives
\begin{equation}
\Br(\BtoROG)=(\BrBtoROG)\EM6, 
\end{equation}
where the first and second errors are statistical and systematic,
respectively.  
The result is consistent with previous
results~\cite{bib:belle-rhogam,bib:babar-rhogam} and in agreement with
SM
predictions~\cite{bib:ali-cdlu,bib:bosch-buchalla}.
The significance of the simultaneous fit is $\signifROratioconv\sigma$,
where the significance is defined as $\sqrt{-2\ln(\Lzero/\Lmax)}$, and
$\Lmax$ ($\Lzero$) is the value of the likelihood function when the
signal branching fraction is floated (set to zero).  Here, the
likelihood function from the fit is convolved with a Gaussian systematic
error function in order to include the systematic uncertainty.
The invariant $\pi\pi(\pi)$ mass and helicity
angle distributions for the events in the signal box are consistent with
those expected from the sum of the signal and background components.
The fit also gives $\Br(\BtoKG)=(\BrBtoKG)\EM6$ (statistical error
only), which is consistent with the world average
value~\cite{bib:hfag2005}.  The individual fit results are in marginal
agreement with the isospin relation.  We test our fitting procedure
using MC simulation and find no statistically significant bias.  We
perform MC pseudo-experiments where events are generated according to
the isospin relation; from the two-dimensional distribution of the
deviation between the $B^-$ and averaged $\Bbar^0$ rates and that
between the $\rho^0\gamma$ and $\omega\gamma$ rates, we find the
probability to observe an isospin violation equal to or larger than our
measurement to be 4.9\%.  The expected level of isospin violation is
within $\pm10\%$~\cite{bib:ali-cdlu}.

The systematic error is estimated by varying each of the fixed
parameters by $\pm1\sigma$ and then taking the quadratic sum of the
deviations in the branching fraction from the nominal value.
We note that the ARGUS background shape in the fit to the $\omegaG$ mode
is steeper than those for the other two modes.  Therefore
we also vary the ARGUS shape parameter for the $\omega\gamma$ mode by
$-2\sigma$ and include the deviation in the systematic error.

The ratio ${\Br(\BtoROG)/\Br(\BtoKG)}=\ratioBtoROGfull$, which we obtain
from a separate fit, can be used to determine $|\Vtd/\Vts|$.  The fit
takes into account the correlated systematic errors between the signal
and $\KstarG$ modes and thus gives a reduced total error.  Using the
relation~\cite{bib:ali-2004}
$
 {\Br(\BtoROG)\over\Br(\BtoKG)}=
 \left| {\Vtd\over\Vts} \right|^2
 {(1-m_{(\rho,\omega)}^2/m_B^2)^3 \over (1-m_{K^*}^2/m_B^2)^3}
 \zeta^2
 [1 + \Delta R]
$,
where the form factor ratio $\zeta=0.85\pm0.10$ and the $SU(3)$-breaking
correction $\Delta R=0.1\pm0.1$, we obtain
\begin{equation}
 |\Vtd/\Vts|=\VtdoverVtsValFull.
\end{equation}
We obtain a 95\% confidence level interval of
$\VtdoverVtsLo<|\Vtd/\Vts|<\VtdoverVtsHi$ using an ensemble of MC
samples in which the experimental error is a quadratic sum of the
asymmetric Gaussian statistical and systematic errors, and the theory
error is a flat distribution in the given range.  This result is in
agreement with the range favored by a fit to the unitarity
triangle~\cite{bib:pdg2004} assuming $|\Vts|=|\Vcb|$.


In conclusion, we observe the process $\btodgamma$ using the
$B\to\rho\gamma$ and $\omega\gamma$ modes.  The resulting branching
fractions are consistent with SM
predictions~\cite{bib:ali-cdlu,bib:bosch-buchalla}.
The ratio of the $\BtoROG$ branching fraction to that for $\BtoKG$ is
used to determine $|\Vtd/\Vts|$.


We thank the KEKB group for the excellent operation of the
accelerator, the KEK cryogenics group for the efficient operation of
the solenoid, and the KEK computer group and the NII for valuable
computing and Super-SINET network support.  We acknowledge support
from MEXT and JSPS (Japan); ARC and DEST (Australia); NSFC (contract
No.~10175071, China); DST (India); the BK21 program of MOEHRD and the
CHEP SRC program of KOSEF (Korea); KBN (contract No.~2P03B 01324,
Poland); MIST (Russia); MHEST (Slovenia); SNSF (Switzerland); NSC and
MOE (Taiwan); and DOE (USA).


\end{document}